\documentclass{ws-procs961x669}            
\usepackage{braket}
\begin{document}
\title{On the semiclassical and quantum picture \\ of the Bianchi I polymer dynamics}

\author{E. Giovannetti$^*$}

\address{Department of Physics, La Sapienza University of Rome, Rome, Italy\\
$^*$Speaker. E-mail: eleonora.giovannetti@uniroma1.it}

\author{G. Montani}

\address{Department of Physics, La Sapienza University of Rome, Rome, Italy\\
Fusion and Nuclear Safety Department, ENEA, Frascati (RM), Italy}

\author{S. Schiattarella}

\address{Department of Physics, La Sapienza University of Rome, Rome, Italy}

\begin{abstract}
We analyze the Bianchi I cosmology in the presence of a massless scalar
field and describe its dynamics via a semiclassical and quantum polymer
approach. We investigate the morphology of the emerging Big Bounce
by adopting three different sets of configurational variables:
the natural Ashtekar connections, the Universe volume plus two anisotropy
coordinates and a set of anisotropic volume-like coordinates
(the latter two sets of variables would coincide in the case of an
isotropic Universe). In the semiclassical analysis we demonstrate that the Big Bounce emerges in the dynamics for all the three sets of variables. Moreover, when the Universe volume itself is considered as a configurational variable, we have derived the polymer-modified Friedmann equation and demonstrated that the Big Bounce has a universal nature, {\it i.e.} the total critical energy density has a maximum value fixed by fundamental constants and the Immirzi parameter only. From a pure quantum point of view, we investigate the Bianchi I dynamics only in terms of the Ashtekar connections. In particular, we apply the Arnowitt--Deser--Misner (ADM) reduction of the variational principle and then we quantize the system. We study the resulting Schr\"{o}dinger dynamics, stressing that the wave packet peak behavior over time singles out common features with the semiclassical trajectories.
\end{abstract}


\bodymatter

\section{INTRODUCTION}

The emergence of a Big Bounce in the quantum dynamics of the isotropic Universe is one of the most relevant result achieved in Loop Quantum Cosmology (LQC) as a phenomenological implication of Loop Quantum
Gravity (LQG) \cite{Rovelli,CQG} when applied to the cosmological sector \cite{ashtekar2003,ashtekar2005gravity,Ashtekar2006,AshtekarI,Ashtekar2008,Ashtekar2011,B}. Although, many criticisms have been expressed towards the LQC framework, especially regarding the lack of a real quantum description of the Big Bounce properties, the unclear equivalence between  the so-called improved scheme with the original LQC formulation in the Ashtekar variables and the process through which the symmetry reduction is performed in order to properly derive LQC as the cosmological sector of LQG \cite{Bojowald_2020}. In this context, Polymer Quantum Mechanics (PQM) \cite{ASHTEKARpol,Pol} represents a convenient mathematical instrument through which deeply analyze bouncing cosmologies, since it is able to reproduce LQC results at least at an effective level and without entering the issues of LQG and LQC. In this work we apply PQM to the Bianchi I model in the presence of a massless scalar field and we explore both its semiclassical and quantum dynamics in different sets of configurational variables \cite{Pol}, in particular the natural Ashtekar connections and two different sets of volume-like variables (the anisotropic volume-like coordinates and the
Universe volume itself plus two anisotropies),
following the LQC formulation in \cite{ashtekar2009,szulc}.

The paper is structured as follows. In Sec. \ref{pol} the theory of PQM is introduced, while the original part of the paper is developed in Secs. \ref{sem1}--\ref{quant}. In particular, in Secs. \ref{sem1} and \ref{sem2} it is solved the semiclassical polymer dynamics of the Bianchi I model in the Ashtekar variables and in the two sets of volume-like variables respectively, then in Sec. \ref{quant} a full quantum treatment in the Ashtekar variables is developed using a Scr\"{o}dinger-like formalism. Finally, in Sec. \ref{con} some concluding remarks are commented.

\section{POLYMER QUANTUM MECHANICS\label{pol}}
PQM is a non-equivalent representation of quantum mechanics with respect to the standard Schrödinger one\cite{Pol} and is based on the assumption that one or more variables of the phase space are discretized.

Firstly, we consider a set of abstract kets $\ket \mu$ with $\mu \in {\rm I\!R}$ with the inner product $\bra \mu \nu\rangle =\delta_{\mu\nu}$, where $\delta_{\mu\nu}$ is a Kronecker delta. This procedure defines the non-separable Hilbert space $H_{poly}$ where two fundamental operators can be introduced: the \emph{label} operator $\hat\epsilon$, whose action on the kets is given by $\hat\epsilon\ket\mu=\mu\ket\mu$, and the \emph{shift} operator $\hat s(\lambda)$ ($\lambda\in {\rm I\!R}$), where $\hat s (\lambda)=\ket{\mu+\lambda}$. The action of $\hat s (\lambda)$ is discontinuous since the kets are orthonormal $\forall \lambda$. In this sense, the set of kets indexed by $\mu$ is said to be discrete.

Now, we consider a one-dimensional system identified by the coordinates $(q,p)$ and we suppose that the coordinate $q$ has a discrete character. In the $p$-polarization the wave function writes as ($\hbar=1$) 	$\psi_{\mu}(p)=\langle p|\mu\rangle=e^{ip\mu}$, the action of the operator $\hat q$ is differential and in particular coincident with the label operator. Its eigenvalues are precisely the $\mu$ parameters and constitute a discrete set. On the other hand, in this polarization the shift operator acts in a multiplicative way as
\begin{equation}
	\hat s(\lambda)\psi_{\mu}(p)=e^{i\lambda p}e^{i\mu p}=\psi_{(\mu+\lambda)}(p)\,.
\end{equation}
As already mentioned, its action is discontinuous; therefore, we must conclude that the operator $\hat p$ does not exist since it is not possible to obtain a discontinuous operator from the exponentiation of a Hermitian one. Similar conclusions are reached by analyzing the structure of the $q$-polarization. Indeed, defining the variable $q$ as discrete implies the non-existence of the operator $\hat{p}$, which must be regularized in order to deal with a well-defined dynamics. In order to overcome this issue, we introduce a regular graph
$\gamma_{\mu_{0}}=\{q\in{\rm I\!R}\,|\, q=n\mu_{0}, \forall n \in \mathbb{Z}\}$, {\it i.e.} a numerable set of equidistant points whose spacing is given by the scale $\mu_{0}$, and we restrict the action of the shift operator $e^{i\lambda p}$ by imposing $\lambda=n \mu_{0}$ in order to remain in the lattice. Its action is well-defined, so we use it to approximate any function of $p$ in the following way:
\begin{equation}
	\label{approxPol}
	p \approx \frac{1}{\mu_{0}} \sin \left(\mu_{0} p\right)=\frac{1}{2 i \mu_{0}}\left(e^{i \mu_{0} p}-e^{-i \mu_{0} p}\right)\,.
\end{equation}
We notice that this approximation is good for $\mu_{0}p\ll1$. Under this hypothesis, the action of the regularized $\hat p$ operator is
\begin{equation}
	\label{phat}
	\hat{p}_{\mu_{0}}\ket{\mu_{n}}=-{i\over 2\mu_{0}}(\ket{\mu_{n+1}}-\ket{\mu_{n-1}})\,.
\end{equation}
In the original part of the paper we will apply this picture to the configurational variables describing the Bianchi I cosmology.
The interest we have in using a representation of quantum mechanics characterized by a discrete configurational variable relies in investigating cut-off effects on geometrical quantities and hence on the cosmological dynamics.

\section{SEMICLASSICAL POLYMER DYNAMICS OF THE BIANCHI I MODEL IN THE ASHTEKAR VARIABLES\label{sem1}}

In this section we firstly investigate the semiclassical polymer dynamics of the Bianchi I model in terms of the Ashtekar variables. The phase space is six-dimensional and it is expressed through the canonical couple $(c_{i},p_j)$, which link with the scale factors $a_i$ and their velocities is
\begin{equation}
	p_{i}=|\epsilon_{ijk}a_{j}a_{k}|\text{sign}(a_{i})\,, \hspace{0.5cm}c_{i}=\gamma\dot a_{i}\,,
\end{equation}
with $i=1,2,3$ and $\{c_{i},p_j\}=\kappa\gamma\delta_{ij}$. In  order to implement the polymer paradigm, we proceed by imposing the polymer substitution for the Ashtekar connections		
\begin{equation}
	c_i\to{1\over\mu_i}sin({\mu_ic_i})\,,
\end{equation}
so the polymer Hamiltonian constraint with a massless scalar field takes the form
\begin{align}
	\label{eqn:Hpoly}
	\mathcal{H}_{poly}=&-{1\over{\kappa\gamma^2V}}\sum_{i\neq j}{{\sin(\mu_ic_i)p_i\sin(\mu_jc_j)p_j}\over\mu_i\mu_j}+{p^2_{\phi}\over 2V}=0\,,
\end{align}
where $i,j=1,2,3$ and $V=\sqrt{p_1p_2p_3}$. This way, we have chosen to define the variables $p_i$ on three independent polymer lattices (with spacing $\mu_i$ respectively) due to their geometrical character. We also choose $\phi$ as the internal time, so $N$ is fixed by the gauge $N={{\sqrt{p_1p_2p_3}}\over p_\phi}$, so that the equations of motion in the polymer representation are:
\begin{equation}
	\label{Bsystem}
	\begin{cases}
		\begin{aligned}
			\small &{{dp_i}\over{d\phi}}=-{{p_i\cos(\mu_ic_i)}\over\gamma p_{\phi}}\Big[{p_j\over{\mu_j}}\sin(\mu_jc_j)+{p_k\over{\mu_k}}\sin(\mu_kc_k)\Big]\\
			\small &{{dc_i}\over{d\phi}}={\sin(\mu_ic_i)\over{\gamma\mu_i p_{\phi}}}\Big[{p_j\over{\mu_j}}\sin(\mu_jc_j)+{p_k\over{\mu_k}}\sin(\mu_kc_k)\Big]
		\end{aligned}
	\end{cases}
\end{equation}
for $i,j,k=1,2,3$, $i\neq j\neq k$. 
\begin{figure}[h!]
	\centering
	\includegraphics[width=0.55\linewidth]{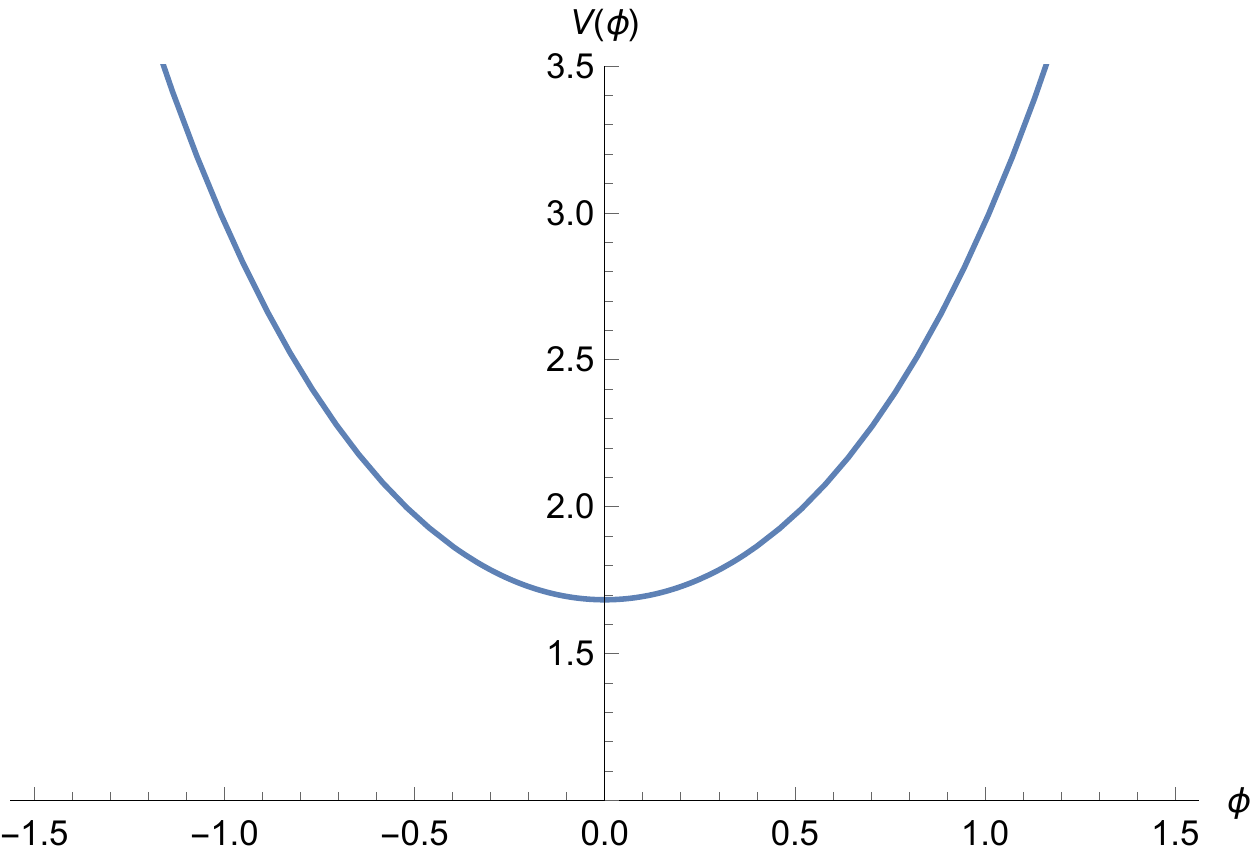}
	\caption{Polymer trajectory of the Universe volume $V=\sqrt{p_{1}p_{2}p_{3}}$ in function of time $\phi$: the Big Bounce replaces the initial singularity of the Bianchi I model.}
	\label{V(phi)}
\end{figure}
By identifying proper constants of motion, the equations in \eqref{Bsystem} can be decoupled and the system can be analytically solved. In particular, in Fig. \ref{V(phi)} it is shown the behavior of the Universe volume $V(\phi)=\sqrt{p_1(\phi)p_2(\phi)p_3(\phi)}$ as function of $\phi$. The resulting trajectory highlights that a semiclassical Big Bounce replaces the classical Big Bang in the polymer framework.

\section{SEMICLASSICAL POLYMER DYNAMICS OF THE BIANCHI I MODEL IN THE VOLUME-LIKE VARIABLES \label{sem2}}

In this section we study the dynamics of the Bianchi I Universe for a new choice of variables. In particular, the anisotropic character of the Bianchi I model leads to the possibility of considering two different sets of volume-like variables, that coincide in the case of the isotropic model.

\subsection{Analysis in the anisotropic volume-like variables: $(V_{1},V_{2},V_{3})$}

Firstly, we consider three equivalent generalized coordinates which correspond to the proper Universe volume in the isotropic limit only \cite{szulc}:
\begin{equation}
	V_{i}=\text{sign}(p_{i})|p_{i}|^{3\over2}\,,\hspace{0.5cm}\eta_{i}={2c_{i}\over3\sqrt{|p_{i}|}}\,,
\end{equation}
where $\eta_i$ for $i=1,2,3$ are the conjugate momenta and the new symplectic structure for the system is characterized by the Poisson brackets $\{\eta_{i},V_{j}\}=\kappa\gamma\delta_{ij}$.

After that the canonical transformation has been performed, we obtain the polymer Hamiltonian constraint in the new variables by using the polymer substitution for the momenta $\eta_i$, since the variables $V_i$ live on the corresponding polymer lattices. It reads as
\begin{equation}
	\label{Hv}
	\mathcal{H}_{poly}=-{9\over{4\kappa\gamma^{2}V}}\sum_{i\neq j}{V_{i}\sin(\mu_{i}\eta_{i})V_{j}\sin(\mu_{j}\eta_{j})\over\mu_{i}\mu_{j}}+{p_{\phi}^{2}\over 2 V}=0\,,
\end{equation}
where $i,j=1,2,3$ and $V=(V_{1}V_{2}V_{3})^{1\over3}$. Analogously, we impose $N=\frac{V}{p_\phi}$ due to the choice of $\phi$ as relational time, so the Hamilton equations describing the dynamics are
\begin{equation}
	\label{sysV}
	\begin{cases}
		\begin{aligned}
			&\frac{dV_i}{d\phi}=-{{9V_i\cos(\mu_i\eta_i)}\over 4\gamma p_{\phi}}\bigg[{V_j\over{\mu_j}}\sin(\mu_j\eta_j)+{V_k\over{\mu_k}}\sin(\mu_k\eta_k)\bigg]\\
			&\frac{d\eta_i}{d\phi}={9\sin(\mu_i\eta_i)\over{4\gamma\mu_1p_{\phi}}}\bigg[{V_j\over{\mu_j}}\sin(\mu_j\eta_j)+{V_k\over{\mu_k}}\sin(\mu_k\eta_k)\bigg]
		\end{aligned}
	\end{cases}
\end{equation}
for $i\neq j\neq k$. In analogy with the previous treatment, we can identify proper constants of motion that decouple the system \eqref{sysV}  along the three directions. 

\begin{figure}[h!]
	\centering
	\includegraphics[width=0.55\linewidth]{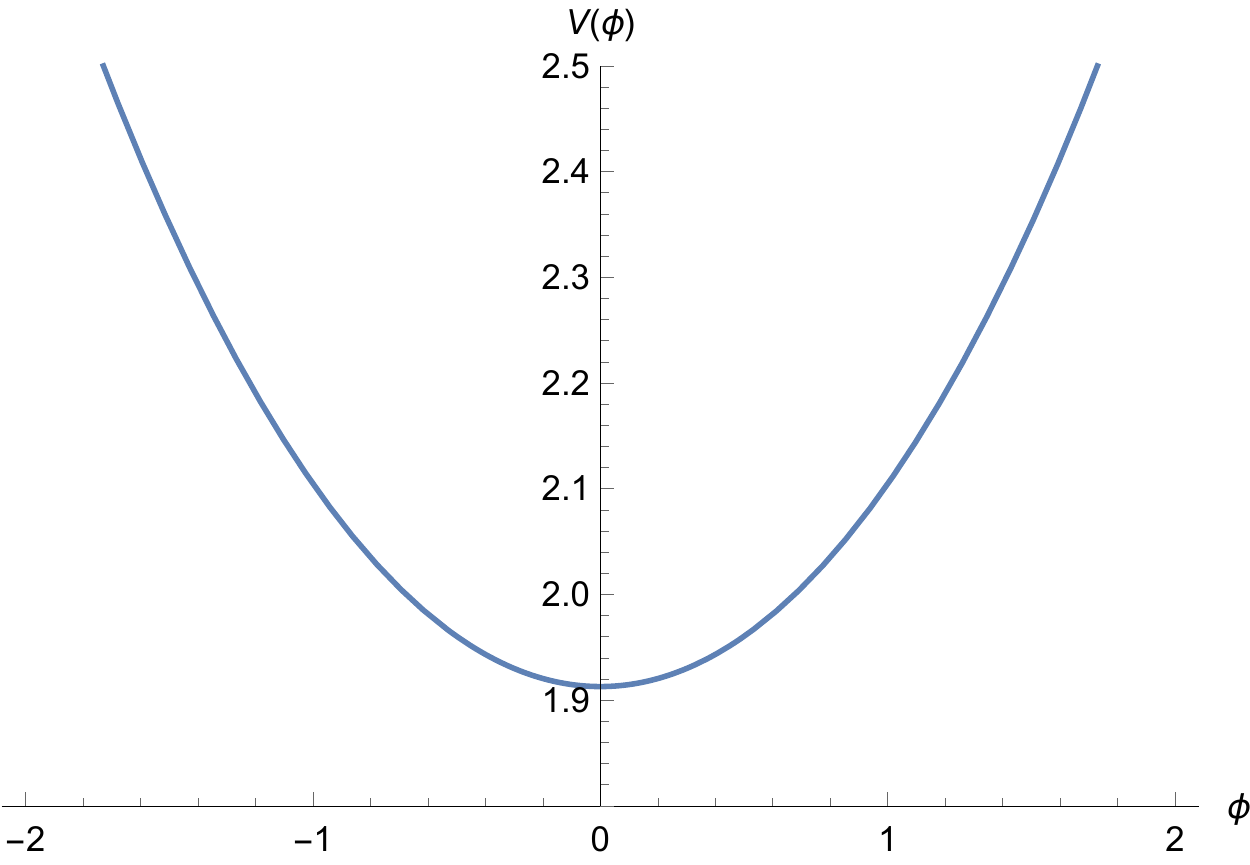}
	\caption{Semiclassical polymer trajectory of the Universe volume $V=(V_1V_2V_3)^{1/3}$ as function of $\phi$.}
	\label{Vv}
\end{figure}
By taking general initial conditions according to \eqref{Hv}, analytical solutions for the anisotropic volume coordinates can be gained. In particular, the Universe volume as function of $\phi$ reads as $V(\phi)=(V_1(\phi)V_2(\phi)V_3(\phi))^{1/3}$ and its behavior is shown in Fig. \ref{Vv}. The Big Bounce clearly appears as a polymer regularization effect instead of the classical Big Bang.

\subsection{Analysis in the volume variables: $(v,\lambda_{1},\lambda_{2})$}
\label{SecV}
In the new set of volume variables $(\lambda_{i}=\text{sign}(p_{i})\sqrt{|p_{i}|}\,,\, v=\lambda_{1}\lambda_{2}\lambda_{3})$ the Universe volume itself $v$ is considered and $\eta_{1,2}=2\sqrt{p_{1,2}}c_{1,2}\,,\,\eta_3=2\sqrt{\frac{p_3}{p_1p_2}}c_3$ are the conjugate momenta. The Poisson brackets are conserved ($\{\eta_i,\lambda_j\}=\kappa\gamma\delta_{ij}$ for $i\neq j$, $i,j=1,2$ and $\{\eta_3,v\}=\kappa\gamma$) and the semiclassical polymer Hamiltonian takes the form
\begin{equation}
		\label{HV}
		\mathcal{H}_{poly}=-{1\over{4\kappa\gamma^{2}V}}\Big(\sum_{i=1,2}{\lambda_{i}\sin(\mu_{i}\eta_{i})v\sin(\mu_{3}\eta_{3})\over\mu_{i}\mu_{3}}+{\lambda_{1}\sin(\mu_{1}\eta_{1})\lambda_{2}\sin(\mu_{2}\eta_{2})\over\mu_{1}\mu_{2}}\Big)+{p_{\phi}^{2}\over 2 V}=0\,,
\end{equation}
where the canonical transformation has been performed before the implementation of the semiclassical polymer paradigm on the variables $\eta_i$, as in the previous subsection.
Analogously, we derive the Hamilton equations for the couple of variables ($v,\eta_{3}$)
\begin{equation}
	\label{vphi}
	\begin{cases}
		\begin{aligned}
			&\frac{dv}{d\phi}=-{{v\cos(\mu_3\eta_3)}\over 4\gamma p_{\phi}}\bigg[{\lambda_1\over{\mu_1}}\sin(\mu_1\eta_1)+{\lambda_{2}\over{\mu_2}}\sin(\mu_2\eta_2)\bigg]\\
			&\frac{d\eta_3}{d\phi}={\sin(\mu_3\eta_3)\over{4\gamma\mu_3p_{\phi}}}\bigg[{\lambda_1\over{\mu_1}}\sin(\mu_1\eta_1)+{\lambda_{2}\over{\mu_2}}\sin(\mu_2\eta_2)\bigg]
		\end{aligned}
	\end{cases}
\end{equation}
and also for the conjugate variables $(\lambda_1,\eta_1)$, $(\lambda_2,\eta_2)$
\begin{equation}
	\begin{cases}
		\begin{aligned}
			&\frac{d\lambda_i}{d\phi}=-{{\lambda_i\cos(\mu_i\eta_i)}\over 4\gamma p_{\phi}}\bigg[{v\over{\mu_3}}\sin(\mu_3\eta_3)+{\lambda_{j}\over{\mu_j}}\sin(\mu_j\eta_j)\bigg]\\
			&\frac{d\eta_i}{d\phi}={\sin(\mu_i\eta_i)\over{4\gamma\mu_ip_{\phi}}}\bigg[{v\over{\mu_3}}\sin(\mu_3\eta_3)+{\lambda_{j}\over{\mu_j}}\sin(\mu_j\eta_j)\bigg]
		\end{aligned}
	\end{cases}
\end{equation}
where we have used
$N=\frac{V}{p_\phi}$
in order to derive the dynamics of the model in function of the relational time $\phi$.

\begin{figure}[h!]
	\centering
	\includegraphics[width=0.5\linewidth]{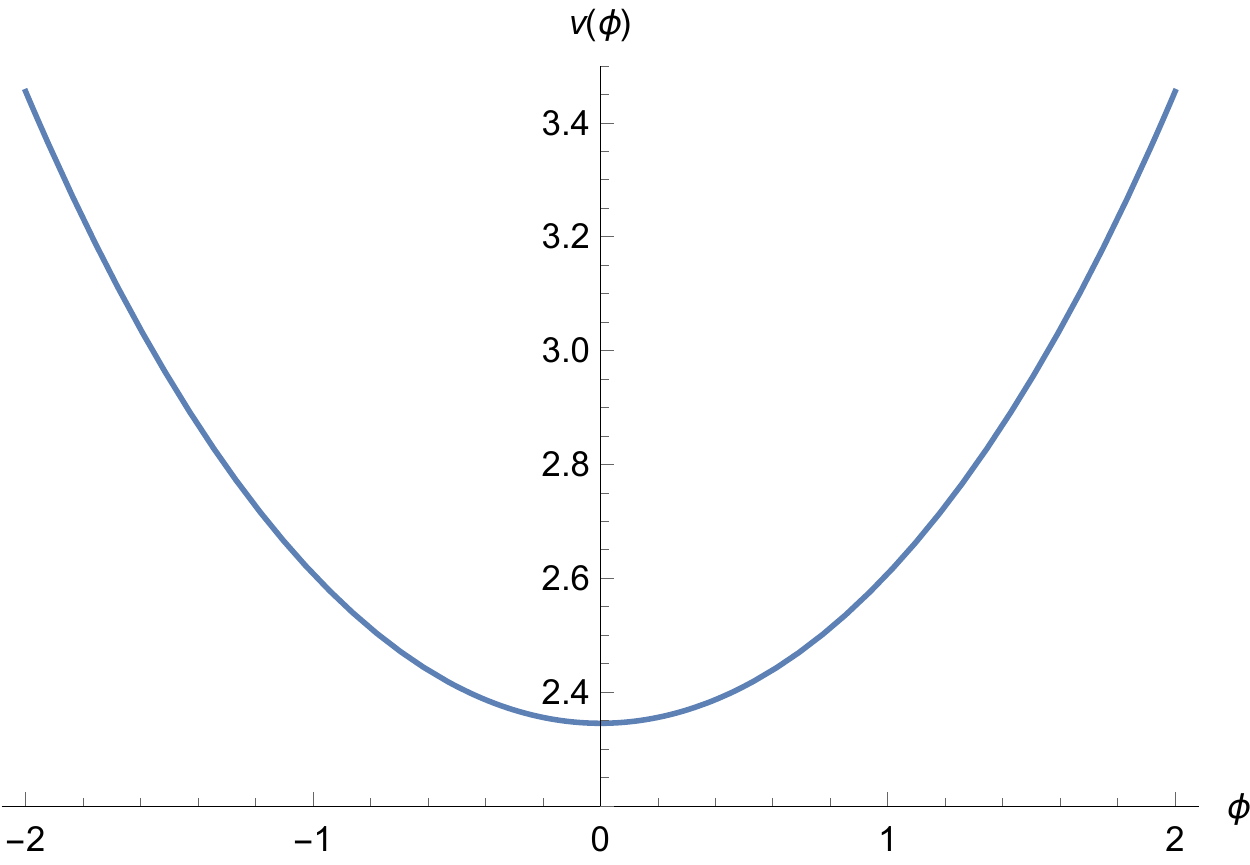}
	\caption{Semiclassical polymer trajectory of the Universe volume $v(\phi)$.}
	\label{vol}
\end{figure}

Once fixed the initial conditions on the variables $(\lambda_1,\eta_1)$, $(\lambda_2,\eta_2)$, $(v,\eta_3)$ according to \eqref{HV}, we can solve this system analytically since the 3D-motion is decoupled in three one-dimensional trajectories, thanks to the use of constants of motion analogous to those ones of the previous cases. In particular, the analytical solution for the Universe volume $v(\phi)$ clearly resembles a bouncing behavior, as shown in Fig. \ref{vol}. 
Also, by choosing a particular shape for these constants of motion, we find the following convenient form for the polymer-modified Friedmann equation of the Bianchi I model in the proper volume variables:

\begin{equation}
	\label{FRIEDMANN}
	H^2=\frac{\kappa}{54}\frac{p_\phi^2+\bar{\mathcal{K}}^2}{v^2}\Big[1-\frac{4\kappa\gamma^2\mu^2}{3}\Big(\frac{p_\phi^2+\bar{\mathcal{K}}^2}{2v^2}\Big)\Big]\,,\;\,\bar{\mathcal{K}}={\mathcal{K}\over{\sqrt{2\kappa\gamma^2\,}}}\,,
\end{equation}
where we have fixed $\mu_1=\mu_2=\mu_3=\mu$ without loss of generality. In \eqref{FRIEDMANN}, the additional term $\bar{\mathcal{K}}^2/2v^2$ reasonably mimics the energy-like contribution of the anisotropies to the total energy density ($\mathcal{K}$ is an arbitrary constant) and the total critical energy density results to be
\begin{equation}
	\rho^{tot}_{crit}=\frac{3}{4\kappa\gamma^2\mu^2}\,,
	\label{tcedv}
\end{equation}
so it has universal features since it is independent from the initial conditions on the motion. By concluding,  thanks to equation \eqref{FRIEDMANN} we have inferred the physical properties of the critical point, showing that taking the Universe volume itself as a configurational variable makes the Big Bounce acquire universal physical properties.

\section{QUANTUM ANALYSIS\label{quant}}

The purpose of this section is studying the Bianchi I model in the Ashtekar variables at a quantum level by applying the Dirac quantization. Firstly, we perform an ADM reduction of the variational principle at a semiclassical level in order to deal with a Schr\"odinger-like formalism. More specifically, after choosing the scalar field $\phi$ as the temporal parameter, we derive the ADM-Hamiltonian by solving the scalar constraint \eqref{eqn:Hpoly} with respect to the momentum associated to the scalar field:
\begin{equation}
	p_{\phi}\equiv\mathcal{H}_{ADM}=\sqrt{\Theta}\,,\quad	\Theta={2\over\kappa\gamma^{2}}\sum_{i\neq j}{{\sin(\mu_ic_i)p_i\sin(\mu_jc_j)p_j}\over\mu_i\mu_j}\,,
	\label{ADM}
	\end{equation}
where $i,j=1,2,3$. We choose the positive root in order to guarantee the positive character of the lapse function (see Sec. \ref{sem1}).
Now we promote the ADM-Hamiltonian to a quantum operator, obtaining the following Schr\"odinger-like equation:
\begin{equation}
	\label{eqn:schro}
	-i\partial_{\phi}\Psi=\sqrt{\hat{\Theta}}\Psi\,,\quad\sqrt{\hat{\Theta}}= \Big[{2\over\kappa^{2}\gamma^{2}}\Big( \partial_{x_{1}}\partial_{x_{2}}+\partial_{x_{1}}\partial_{x_{3}}+ \partial_{x_{2}}\partial_{x_{3}}\Big)\Big]^{1/2}\,,
\end{equation}
where we have used the substitution $x_i=\ln\big[{\tan\big({\frac{\mu c_i}{2}}}\big)\big]+\bar{x}_i$.
The associated probability density is $\mathcal{P}(\vec x,\phi)=\Psi^{*}(\vec x,\phi)\Psi(\vec x,\phi)$, where 
\begin{equation}
	\Psi(\vec{x},\phi)=\int_{-\infty}^{\infty}dk_1\,dk_2\,dk_3\,\prod_{i=1}^3\exp\bigg[-{\frac{(k_i-\mu_{k_i})^2}{2\sigma_{k_i}^2}}\bigg]\,e^{i(k_1x_1+k_2x_2+k_3x_3+\sqrt{2|k_1k_2+k_1k_3+k_2k_3|}\phi)}\,.
	\label{packbis} 
\end{equation}
\begin{figure*}
	\begin{minipage}{3.2cm}
		\includegraphics[width=1.5\linewidth]{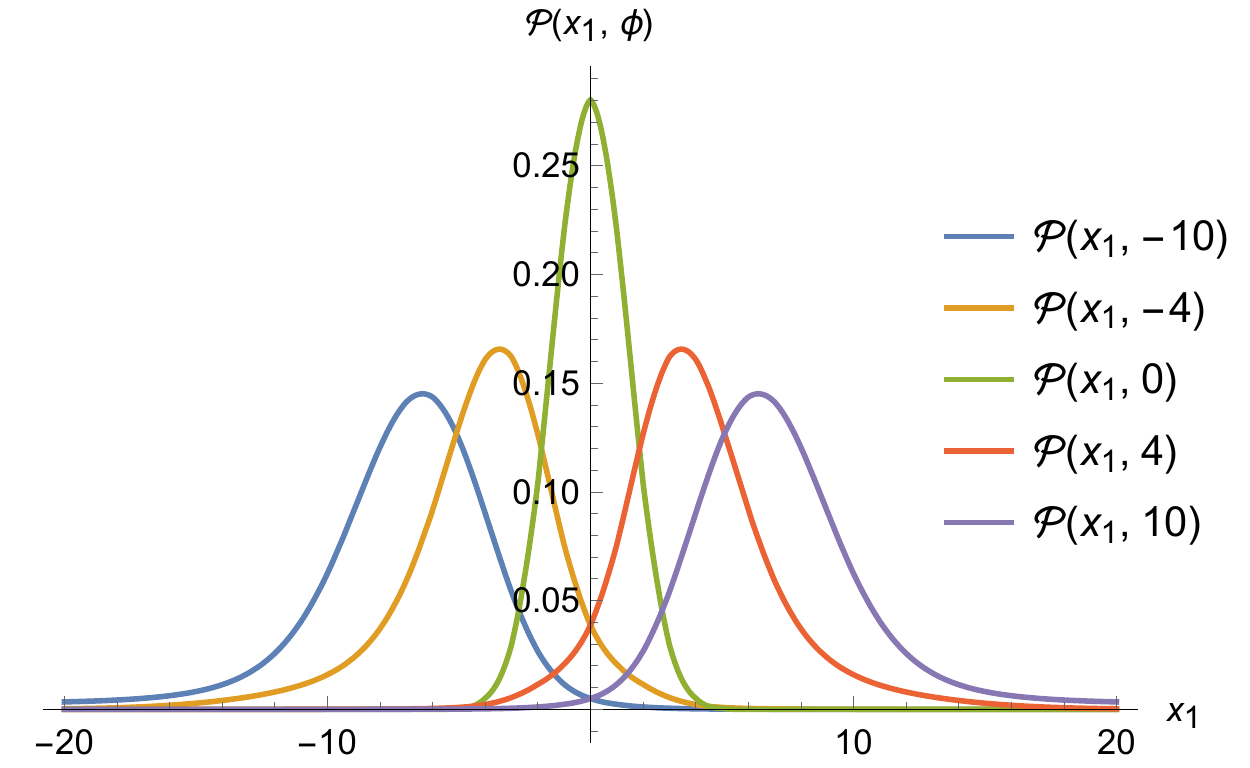}
	\end{minipage}
	\qquad\quad
	\begin{minipage}{3.2cm}
		\includegraphics[width=1.5\linewidth]{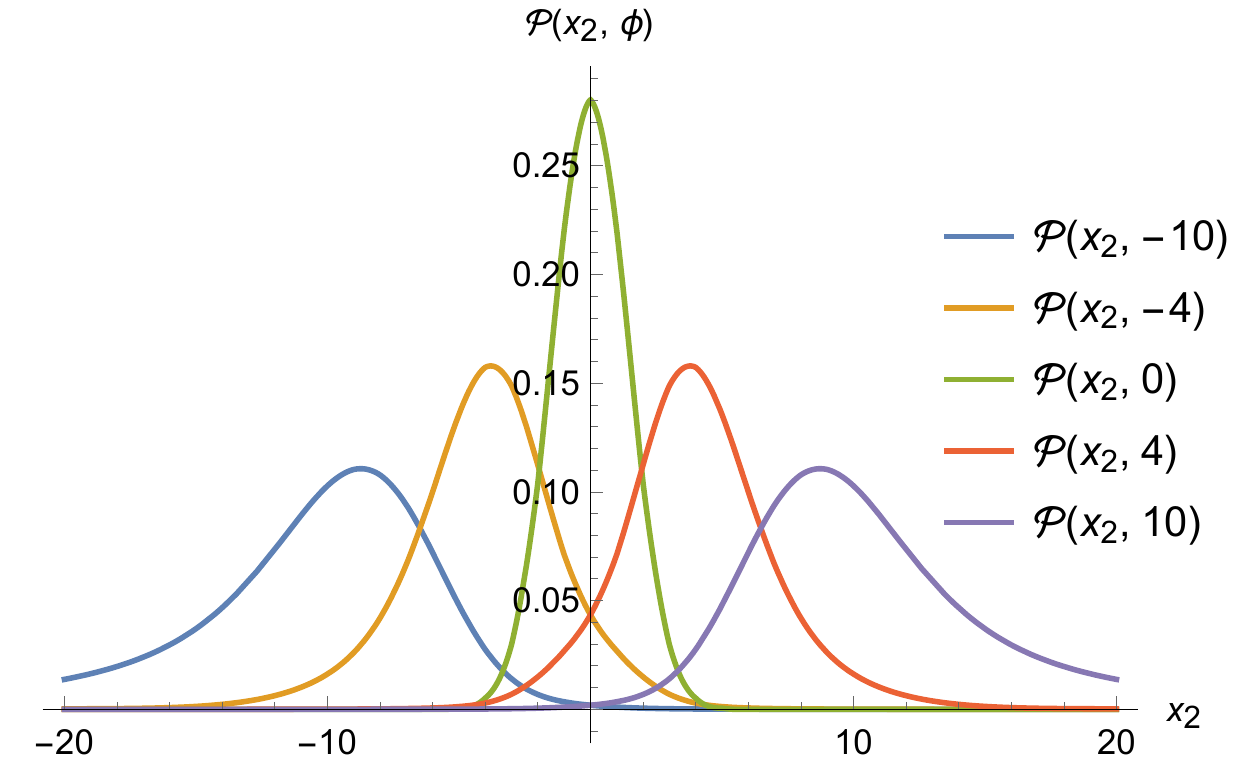}
	\end{minipage}
	\qquad\quad
	\begin{minipage}{3.2cm}
		\includegraphics[width=1.5\linewidth]{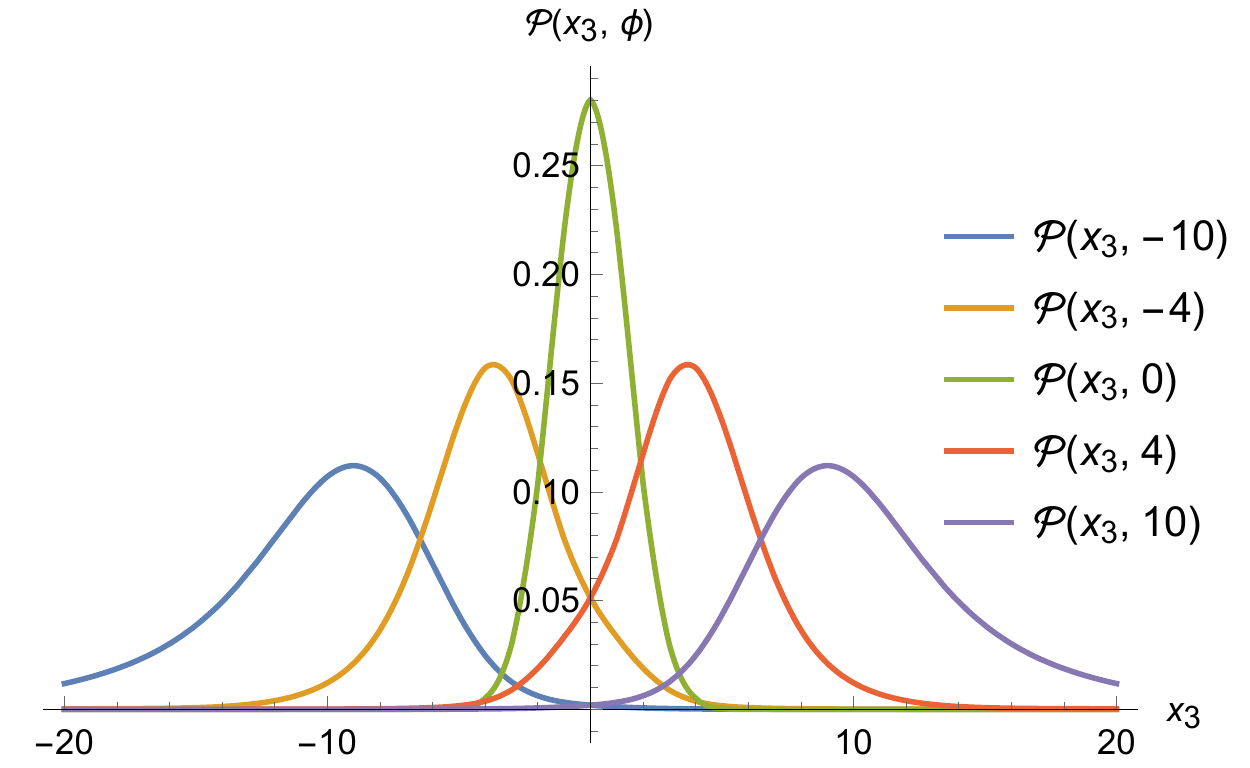}
	\end{minipage}
	\caption{The normalized sections $\mathcal{P}(x_i,\phi)$ are shown in sequence for $i=1,2,3$ respectively at different times (here $\mu_{k_1}=\mu_{k_2}=\mu_{k_3}=0, \sigma_{k_1}=\sigma_{k_2}=\sigma_{k_3}=1/2$). Their spreading behavior over time is evident together with the gaussian-like shape.}
	\label{qprob}
\end{figure*}

\begin{figure*}
	\begin{minipage}{2.8cm}
		\includegraphics[width=1.5\linewidth]{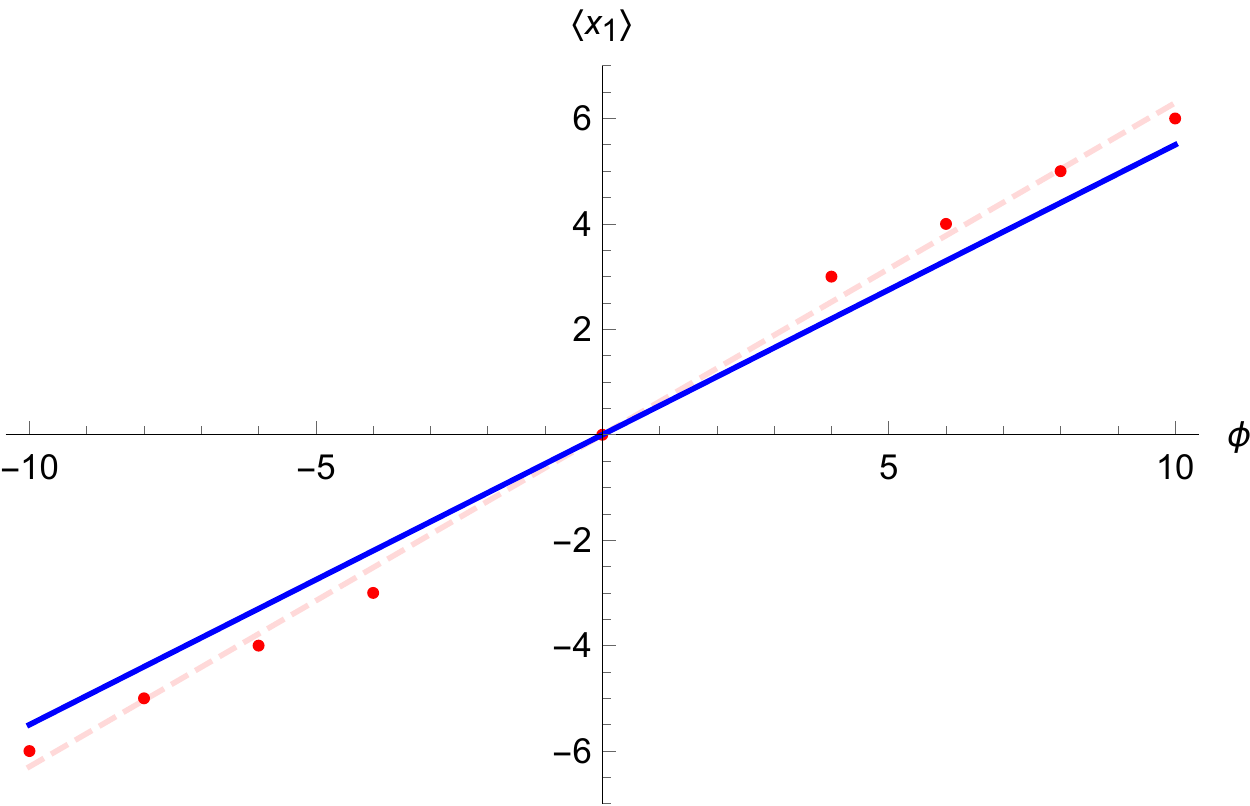}
	\end{minipage}
	\qquad\qquad
	\begin{minipage}{2.8cm}
		\includegraphics[width=1.5\linewidth]{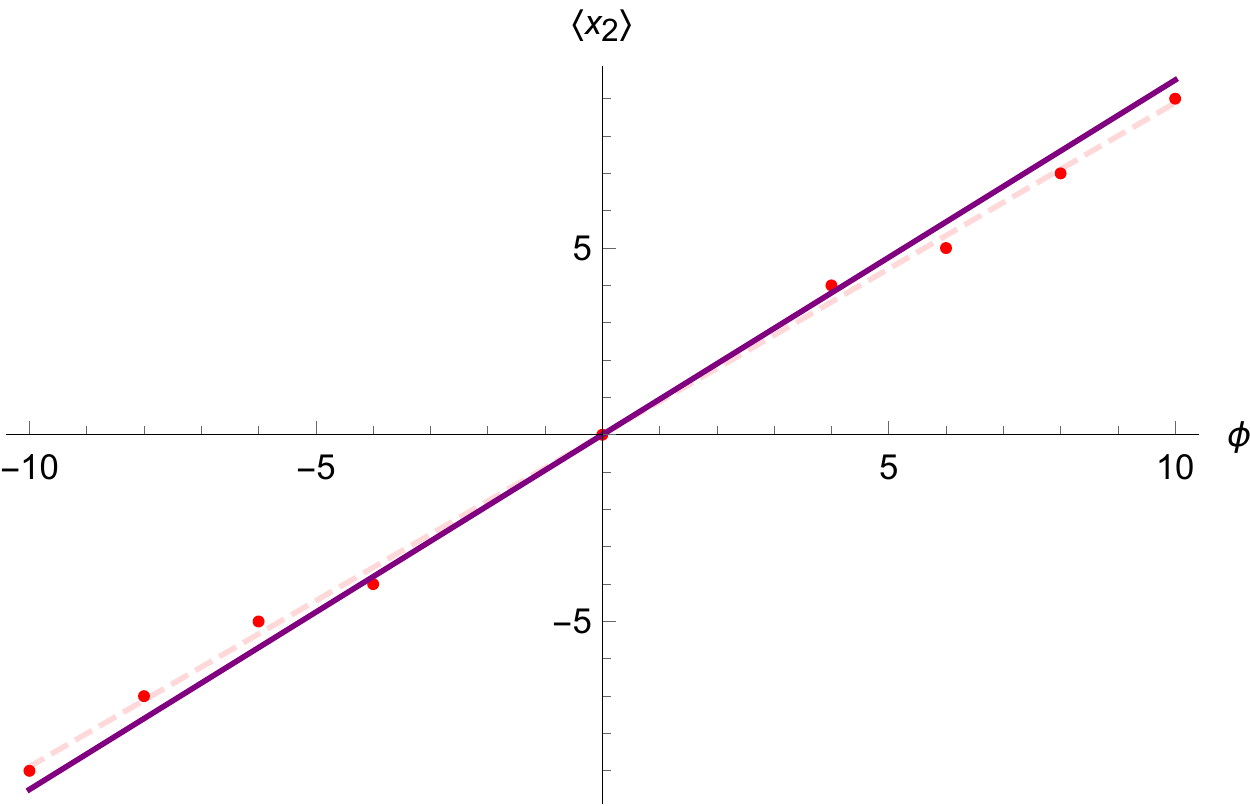}
	\end{minipage}
	\qquad\qquad
	\begin{minipage}{2.8cm}
		\includegraphics[width=1.5\linewidth]{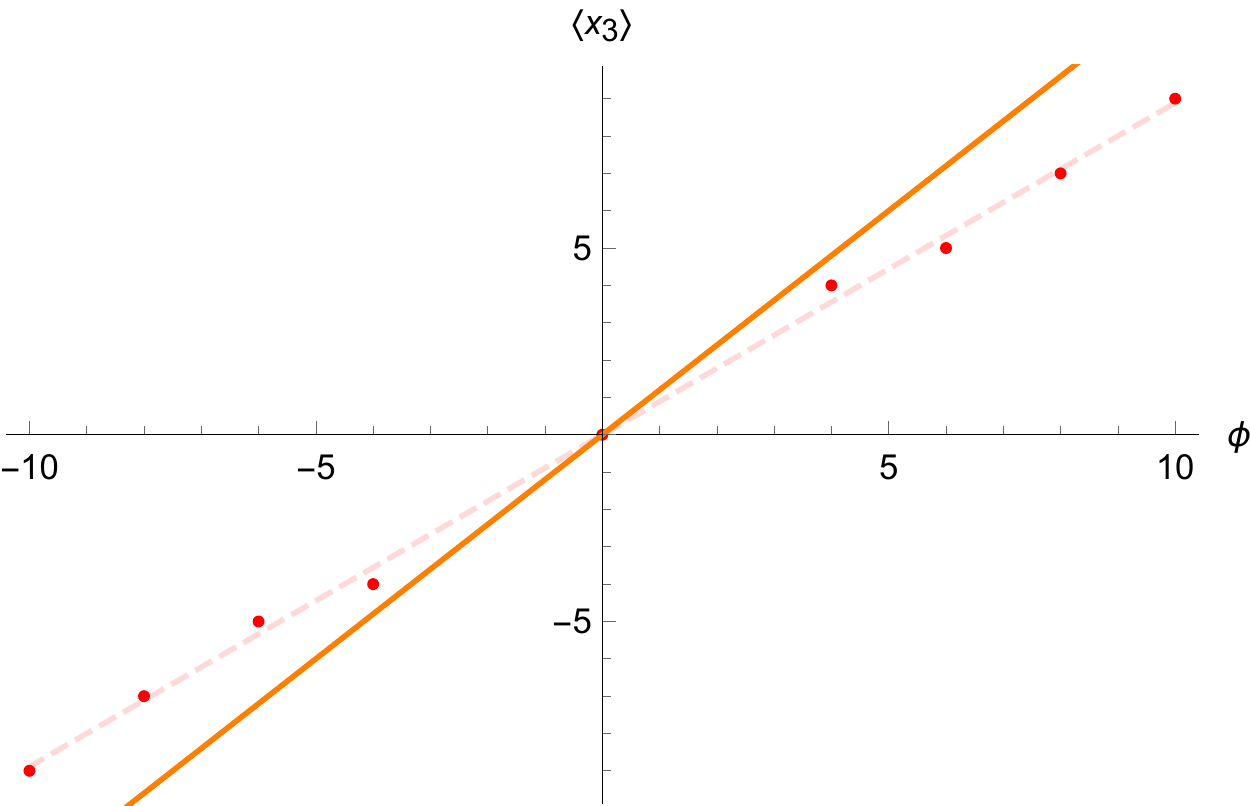}
	\end{minipage}
	\caption{The three pictures show the position of the peaks of $\mathcal{P}(x_i,\phi)$ for $i=1,2,3$ respectively in function of time $\phi$ (red dots). The resulting fitting functions (red dashed straight lines) overlap the semiclassical trajectories (continuous lines) with a confidence level of three standard deviations.}
	\label{qpeak}
\end{figure*}	
Now we can analyze the quantum dynamics of \eqref{packbis} by following the peak of the probability density $\mathcal{P}$ over time, in order to verify the consistency between the information carried by the quantum wave packet of the Bianchi I model and the semiclassical solutions provided in Sec. \ref{sem1} when the Ashtekar variables are considered. In Fig. \ref{qprob} some different sections of the probability density $\mathcal{P}$ at different values of $\phi$ are shown. As we can see, their spreading behavior over time is evident, as well as their gaussian-like shape. Also, in Fig. \ref{qpeak} the position of the peaks of $\mathcal{P}$ are represented by the red dots and have also been fitted by means of a linear interpolation. The resulting fitting functions are represented by the red dashed straight lines, while the semiclassical trajectories by the continuous straight ones. In particular, the slope of the fitting straight lines is consistent with the semiclassical one with a confidence level of three standard deviations for all the three coordinates. Therefore, we conclude that there is a good correspondence between the quantum behavior of the Universe wave function and the solutions of the semiclassical dynamics.

\section{CONCLUDING REMARKS\label{con}}

In this paper we analyzed the polymer semiclassical and quantum dynamics of
the Bianchi I model with a massless scalar field in the Ashtekar variables and in two sets of volume-like ones. In the polymer semiclassical analysis we obtained a bouncing behavior in all the three sets of variables. Also, a polymer-modified Friedmann equation of the Bianchi I model has been derived when the Universe volume itself is considered as a configurational variable. This way, the proper expression of the total critical energy density has been provided, giving a complete picture of the Big Bounce and showing its universal features, {\it i.e.} its dependence from fundamental constants and parameters only. Finally, by performing an ADM reduction of the variational problem, we passed to a Schr\"{o}dinger-like
formalism and then we implemented the canonical quantization in the Ashtekar variables. In particular, the study of the probability density behavior has outlined a good correspondence between the dynamics of the quantum Universe wave packet and the corresponding semiclassical trajectories. In conclusion, due to the role of the Bianchi I model in constructing the general cosmological solution, this study acquires a great relevance in order to understand the properties of the primordial Universe.

\bibliographystyle{ws-procs961x669}
\bibliography{biblio.bib}

\end{document}